%% file: main.tex
\documentclass[smallabstract,smallcaptions]{dccpaper}

\usepackage{fancyhdr}
\usepackage{epsfig}
\usepackage{citesort}
\usepackage{amsmath}
\usepackage{amssymb}
\usepackage{color}
\usepackage{url}
\usepackage{amsmath,amsfonts}
\usepackage{algorithmic}
\usepackage{algorithm}
\usepackage{array}
\usepackage{subfig}
\usepackage{paralist}
\usepackage{url}
\usepackage{textcomp}
\usepackage{stfloats}
\usepackage{verbatim}
\usepackage{graphicx}
\usepackage{cite}
\usepackage{booktabs}
\usepackage{hyperref}
\def \NOTE [#1]{\textcolor{blue}{(\textit{#1})}}
\usepackage{pifont}
\usepackage{bbding}
\usepackage{float}
\usepackage{caption}
\usepackage{graphicx, subfig}
\usepackage{amsthm,amsmath,amssymb}
\usepackage{mathrsfs}
\usepackage{multirow}

\include{macro}

\newlength{\figurewidth}
\newlength{\smallfigurewidth}

\definecolor{cMainSE}{RGB}{201,20,20}
\definecolor{cMainST}{RGB}{201,111,20}
\definecolor{cMainTE}{RGB}{2,119,189}
\begin{document}

\title
{\large
%
\textbf{Stable Diffusion is a Natural Cross-Modal Decoder for Layered AI-generated Image Compression}
}

\author{%
Ruijie Chen$^{\ast}$, Qi Mao$^{\ast}$, Zhengxue Cheng$^{\dag}$\\
{\small\begin{minipage}{\linewidth}\begin{center}
\begin{tabular}{ccc}
    $^{\ast}$Communication University of China & $^{\dag}$Shanghai Jiao Tong University
\end{tabular}
\end{center}\end{minipage}}\\
\begin{tabular}{ccc}
    $\{$chenruijie233, qimao$\}$@cuc.edu.cn & zxcheng@sjtu.edu.cn
\end{tabular}
}

\maketitle
\thispagestyle{empty}

\begin{abstract}
%
Recent advances in Artificial Intelligence Generated Content (AIGC) have garnered significant interest, accompanied by an increasing need to transmit and compress the vast number of AI-generated images (AIGIs).
%
However, there is a noticeable deficiency in research focused on compression methods for AIGIs.
To address this critical gap, we introduce a scalable cross-modal compression framework that incorporates multiple human-comprehensible modalities, designed to efficiently capture and relay essential visual information for AIGIs.
In particular, our framework encodes images into a layered bitstream consisting of a semantic layer that delivers high-level semantic information through text prompts; a structural layer that captures spatial details using edge or skeleton maps; and a texture layer that preserves local textures via a colormap.
Utilizing Stable Diffusion as the backend, the framework effectively leverages these multimodal priors for image generation, effectively functioning as a decoder when these priors are encoded.
Qualitative and quantitative results show that our method proficiently restores both semantic and visual details, competing against baseline approaches at extremely low bitrates ($< 0.02$ bpp). Additionally, our framework facilitates downstream editing applications without requiring full decoding, thereby paving a new direction for future research in AIGI compression.

\end{abstract}


\section{Introduction}
Artificial Intelligence Generated Content (AIGC) refers to the outputs produced by artificial intelligence models especially large multi-modal models (LMMs) that are trained on comprehensive datasets.
%
Recent advancements in AIGC, particularly in the domain of text-to-image (T2I) generation, have been exemplified by popular models such as Stable Diffusion (SD)~\cite{sd}.
%
Unlike Natural Sense Images (NSIs), which are captured by cameras, AI-generated images (AIGIs) can be extensively sampled by T2I models based on provided text descriptions.
Furthermore, controllable generation methods~\cite{controlnet,t2i} integrate T2I models, such as SD, with multiple human-comprehensible prior inputs from a variety of modalities—including edge maps, sketch images, and color palettes—thereby significantly enhancing the controllability of the generated images.
%
%
In the field of image compression, when multiple priors are transmitted to the receiver, the decoding process essentially mirrors the generation process. Within this context, SD serves effectively as a natural cross-modal decoder.

Construction of effective prior representations lays a solid basis for high efficient image compression. 
Such concept, termed Cross-Modal Compression (CMC)~\cite{cmc,vrcmc,rdocmc,rethink}, aims to compress image data into a compact, common, and human-comprehensible domain, such as text, sketches, semantic maps, and other attributions.
However, most existing CMC efforts are limited by their reliance on the narrow capabilities of GAN-based generators\cite{gan} and primarily focus on NSIs. 
 Recently, state-of-the-art methods~\cite{perco,misc,text+sketch} has introduced the use of Large Multimodal Models (LMMs) to enhance image compression. 
Despite their success, they also neglect to extend their application into AIGIs. 
Moreover, current techniques predominantly consider only text prompts for image reconstruction, neglecting other comprehensible modalities that could potentially enhance the process.

In this work, we propose the integration of additional human-comprehensible modalities to develop a scalable cross-modal compression framework specifically tailored for AIGIs. 
Our framework is structured into three distinct layers: the semantic layer, which serves as the foundational layer, and the structure and texture layers, which function as enhancement layers. 
The semantic layer efficiently conveys crucial information in an extremely compact form. 
Subsequently, the enhancement layers improve perceptual quality by incorporating structural and textural information. 
Each layer encodes semantic information in a compact yet comprehensible manner, thereby facilitating downstream applications such as image editing without decoding.
%

The main contributions of this paper can be summarized as follows,

\begin{compactitem}
    \item Our work focuses on AIGI compression, advocating that SD is a natural cross-modal decoder, by effectively leveraging rich and scalable priors information. 
    \item We propose a layered cross-modal compression framework that efficiently encode AIGIs by decomposing priors into semantic, structure and texture layers. The framework captures structural information using either edge or pose representations, depending on the content. 
    \item Qualitative and quantitative results demonstrate the effectiveness of our proposed framework, showing superior performance compared to VVC. Additionally, the framework enables seamless local and global image editing by directly manipulating the scalable bitstream, eliminating the need for full decoding.
\end{compactitem}


\begin{figure*}[!t]
    \centering
    \includegraphics[width=0.98\linewidth]{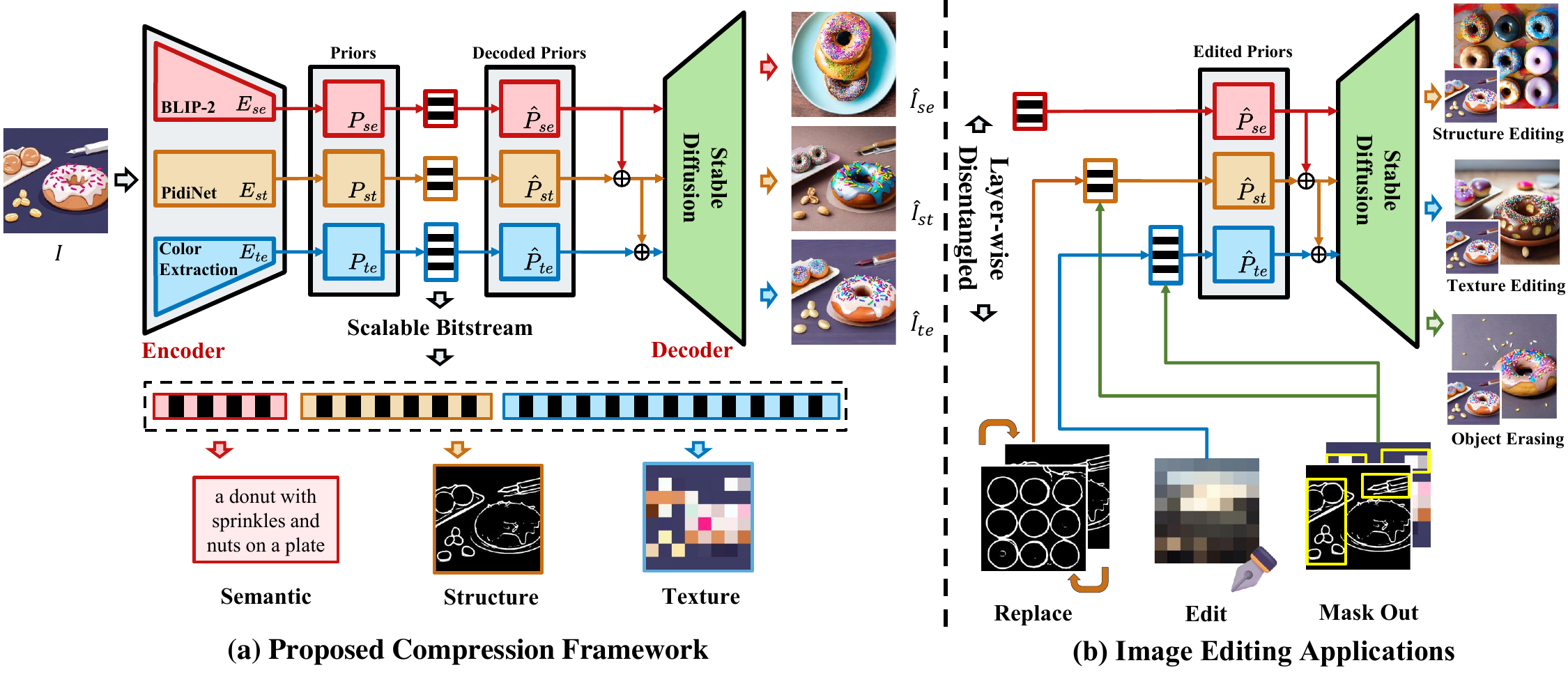}
    \caption{Overview of the proposed layered AGI compression framework using Stable Diffusion as a cross-modal decoder. 
    }
    \label{fig:1-framework}
    \vspace{-5mm}
\end{figure*}

\vspace{-3mm}
\section{Proposed Layered AI-generated Image Compression Framework}
%
%
In this work, we aim to leverage SD~\cite{sd} as a decoder to efficiently compress AIGIs under ultra-low bitrates.  
%
In particular, multi-layered bitstream with semantic, structure, and texture information is constructed as a scalable cross-modal compression framework, tailored to meet varying user requirements for different reconstruction fidelity levels.
As illustrated in \figref{1-framework},  the entire framework consists of three key components: 
%
%
%
%
\begin{compactitem}
    \item Encoder \( E_{\rm{se}} \), \( E_{\rm{st}} \), and \( E_{\rm{te}} \): extract the input image $\mathbf{I}\in \mathbb{R}^{H\times W\times 3}$ into three layered human comprehensible priors.  
    \item Layered priors $P$: are denoted as semantic prior $P_{\rm{se}}$, structure prior $P_{\rm{st}}$ and texture prior $P_{\rm{te}}$, respectively. 
    Each comprehensible prior captures the key visual information for compression, which compensates each other for scalable reconstruction. 
    \item Stable Diffusion as a decoder $D$: reconstruct three-layered images \( \mathbf{\hat{I}}_{\rm{se}}\), \( \mathbf{\hat{I}}_{\rm{st}}\), and \( \mathbf{\hat{I}}_{\rm{te}}\) under the condition of combining different transmitted priors. 
\end{compactitem}

In the following, we detail the
layer-wise priors coding methodologies (\secref{layer-comp}) and the usage of SD~\cite{sd} for
scalable reconstruction (\secref{decoder}) of the proposed framework.

%
%
%

%
\vspace{-3mm}
\subsection{Compressing Layer-wise Priors}
\label{sec:layer-comp}
\vspace{-2mm}
Text prompts, as a key component of the generation of AIGIs, contain key semantic information and guide the selection of visual content from generative models' internal knowledge. 
Therefore, text representations encode high-level semantics in an extremely compact form, serving as the first layer in our framework for preserving key semantic information.
Furthermore, according to Marr's insights~\cite{VISION}, geometric structure (\eg edges) and stochastic texture are two key components that form a visual scene from the perspective of human vision.
Incorporating these two representations helps to mitigate the fidelity gap between images generated solely based on text semantics and original images.
Consequently, on the encoder side, we design three disentangled priors aimed at scalable compression and reconstruction.

\Paragraph{Semantic Prior.}
We obtain the text prompts as semantic prior using commonly-used image-to-text (I2T) model BLIP-2~\cite{blip2} as $E_{\rm{se}}$,
%
\begin{equation}
    P_{\rm{se}} = E_{\rm{se}}(\mathbf{I}).
\end{equation}
%
The extracted text prompts are encoded into bitstream using the Zstd~\cite{Zstd} algorithm for further compression. 
%

%

\Paragraph{Structure Prior.}
Accurately describing geometric structures through natural language is challenging and inefficient. 
As such, the edge map are employed as a structural prior to address this issue.
In particular, we adopt the PidiNet~\cite{pidinet} edge detector as $E_{\rm{st}}$ to extract structure prior,
\begin{equation}\label{eq:eq-st-edge}
  P_{\rm{st}} = E_{\rm{st}}(\mathbf{I}).
\end{equation}
We then down-sample the edge map and compress it via the VVC\cite{vvc} codec. 
Furthermore, for \textbf{human-related} images, human poses can be precisely captured through the combination of body and facial keypoints, forming a pose map that contains only the coordinates and properties of these keypoints—a more compact representation that surpasses edge maps in depicting human geometric structures.
Therefore, we use Openpose~\cite{openpose} as $E_{\rm{st}}$ to extract keypoints as structure prior.
Then, the coordinates are quantized and losslessly compressed into bitstream using the Zstd algorithm~\cite{Zstd}.

\Paragraph{Texture Prior.} 
Texture is another critical component for enhancing reconstruction quality; however, existing image compression methods~\cite{rethink,1k21,humanbody,structureandtexture} implicitly 
 learn texture features, which are not human-comprehensible and limit the flexibility for texture editing.
Hence, we utilize $8\times8$ color palette map to represent the texture of the input image as,
\begin{equation}\label{eq:eq-st-te}
  P_{\rm{te}} = E_{\rm{te}}(\mathbf{I}), 
\end{equation}
%
where $E_{\rm{te}}$ denotes the downsampling operation of the input image $\mathbf{I}$ by a factor of 64.
%

\vspace{-3mm}
\subsection{Using Stable Diffusion as a Decoder}
\label{sec:decoder}
The decoder is designed to reconstruct images using decoded layered priors in \secref{layer-comp} as guidance. 
As such, we leverage the pre-trained SD~\cite{sd} as our universal cross-modal decoder, which allows for scalable image reconstruction across various fidelity levels.

As shown in \figref{1-framework}, 
when only the first layer of semantic prior is transmitted, we expect the partially decoded image from \textcolor{cMainSE}{\textbf{the semantic layer}} $\mathbf{\hat{I}}_{\rm{se}}$ to maintain the basic semantics of the original AIGI,
\begin{equation} 
\mathbf{\hat{I}}_{\rm{se}} = D(\hat{P}_{se},0,0).
\end{equation}
With the addition of the second level of structural information, the reconstructed image from \textcolor{cMainST}{\textbf{the structure layer}}  $\mathbf{\hat{I}}_{\rm{st}}$ is able to recover the original image's geometric shape.
%
\begin{equation}
\mathbf{\hat{I}}_{\rm{st}}  = D(\hat{P}_{se},\hat{P}_{st},0).
\end{equation}
Finally, the image from \textcolor{cMainTE}{\textbf{the texture layer}}  $ \mathbf{\hat{I}}_{\rm{te}}$ reconstructed upon full transmission and decoding of all priors, exhibits high perceptual fidelity: 
\begin{equation}
 \mathbf{\hat{I}}_{\rm{te}}= D(\hat{P}_{se},\hat{P}_{st},\hat{P}_{te}).
\end{equation}
To effectively integrate structural and textural information while preserving the correct high-level semantic integrity in the structure and texture layers, we incorporate the T2I-Adapter~\cite{t2i} into SD~\cite{sd}.
Additionally, a significant advantage of our proposed framework is its ability to manipulate the scalable bitstream directly without decoding the image, as demonstrated in \figref{1-framework}(b), which showcases the framework's efficacy in image editing.

%

\begin{figure*}[!t]
    \centering
    \includegraphics[width=0.98\linewidth]{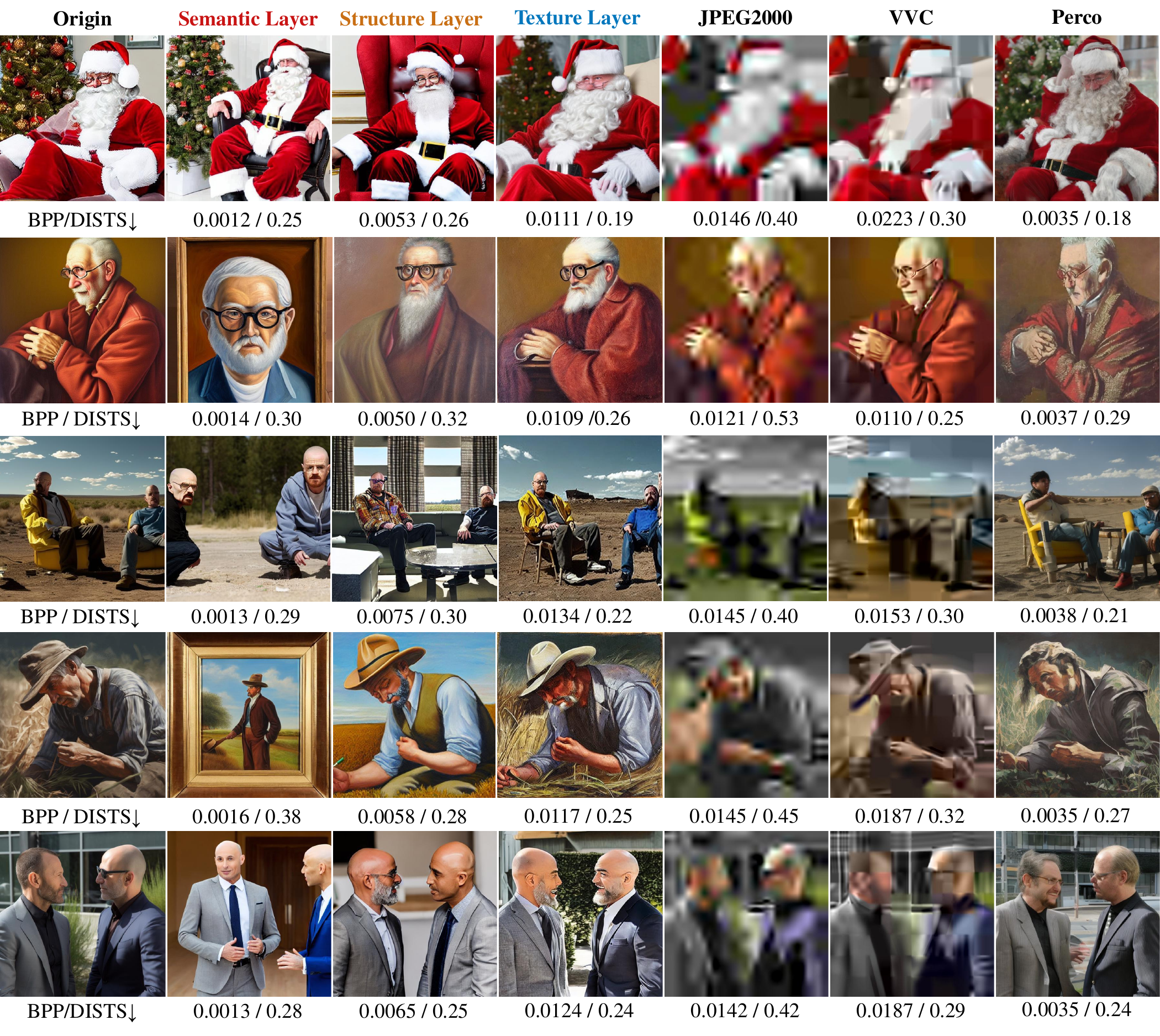}
    \vspace{-3mm}
    \caption{The qualitative comparison results of JPEG2000, VVC, Perco and our method using poses as structure information on the AGIQA dataset.}
    \label{fig:3-pose-qual}
    \vspace{-5mm}
\end{figure*}
%
\vspace{-3mm}
\section{Experimental Results}

\vspace{-3mm}
\subsection{Experimental Setup}
\label{sec:exp-setup}
\vspace{-2mm}
\Paragraph{Datasets.}
We test our proposed compression framework on the AGIQA-3K~\cite{agiqa3k} dataset, which consists $2982$ AIGIs generated by novel text-to-image models with the resolution of $512\times512$. 
%
Images are selected with respect to the two types of structure map designed in \secref{layer-comp}.
For the edge-based type, we select $220$ images in $5$ categories, namely animals, landscape, building, still item, and artwork.
For the pose-based type, we select $162$ images with single human pose detected and $52$ images with multiple poses. 
Pose in each image must occupy no less than $40\%$ of area.
%

\begin{figure*}[!t]
    \centering
    \includegraphics[width=0.98\linewidth]{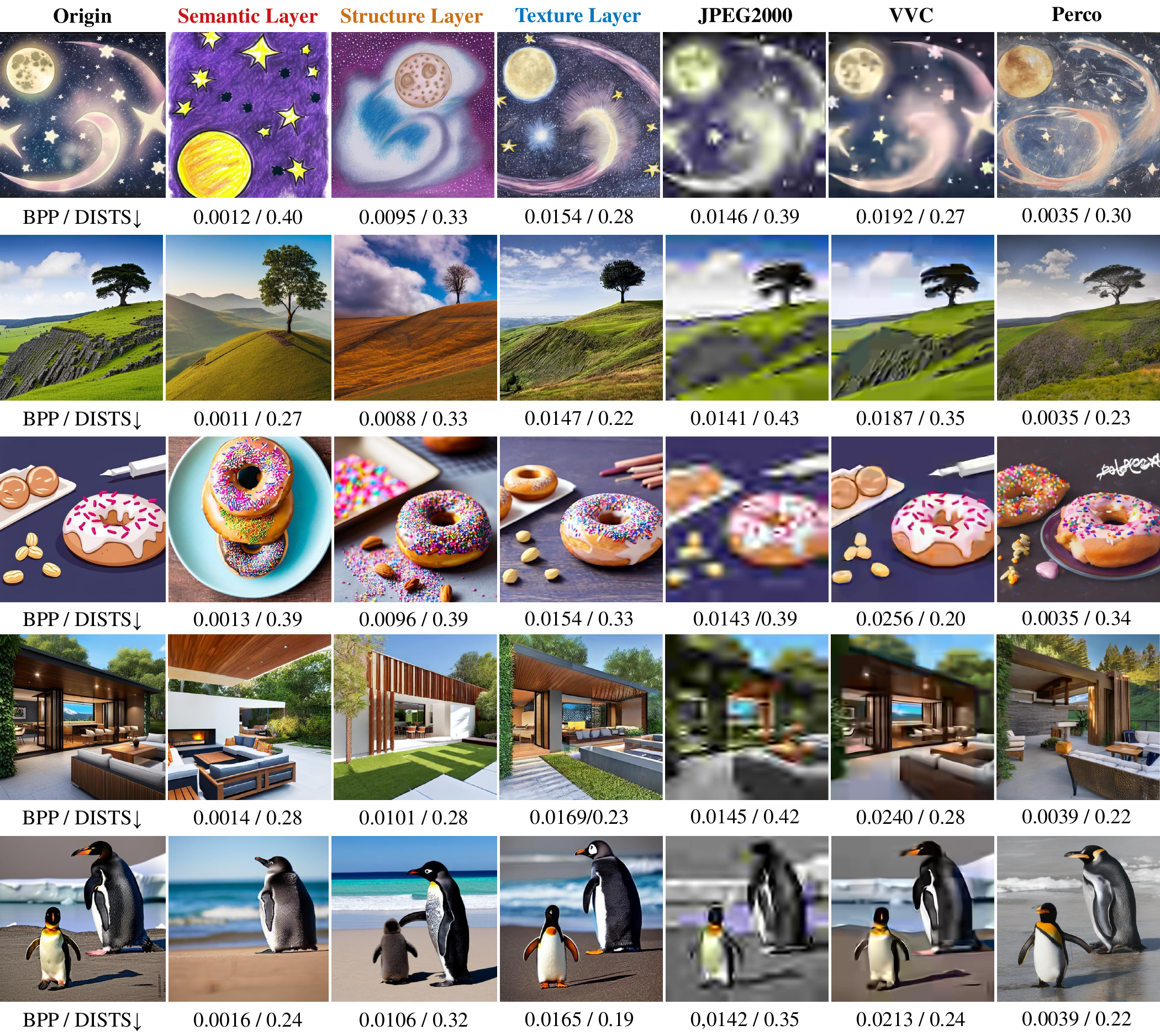}
    \vspace{-3mm}
    \caption{The qualitative comparison results of JPEG2000, VVC, Perco and our method using edges as structure information on the AGIQA dataset.}
    \label{fig:2-edge-qual}
    \vspace{-5mm}
\end{figure*}
\begin{figure*}[!t]
    \centering
    \includegraphics[width=0.98\linewidth]{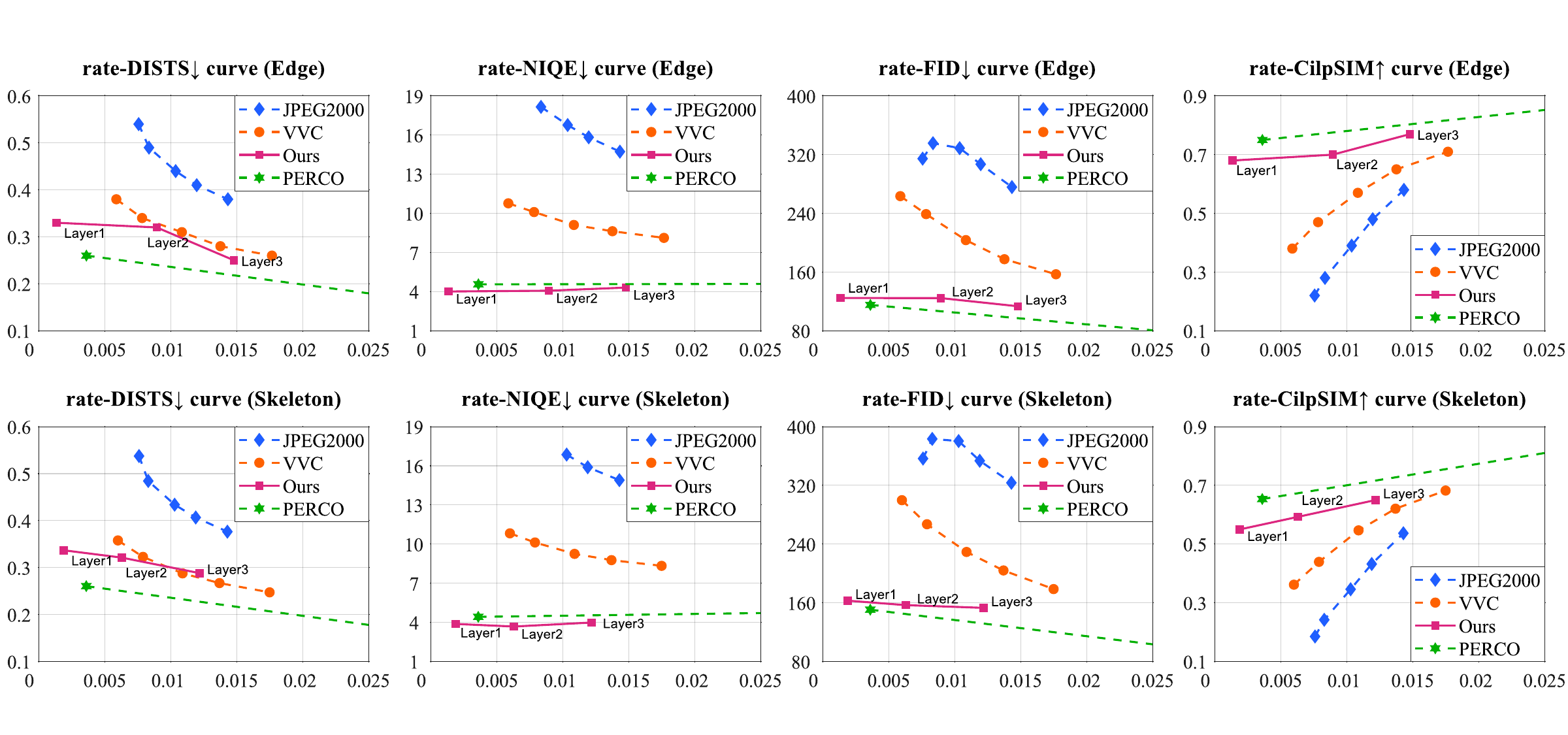}
    \vspace{-5mm}
    \caption{The R-D performance of JPEG2000, VVC, Perco and our method on the AGIQA dataset. Layer1, Layer2 and Layer3 denotes images incrementally reconstructed from semantic, structure and texturte priors, respectively.}
    \label{fig:4-quant}
    \vspace{-5mm}
\end{figure*}
%
\Paragraph{Evaluation Metrics.}
In the proposed scheme, we aim to preserve semantic and perceptual quality. 
To evaluate the semantic fidelity, we utilize Fréchet Inception Distance (FID)~\cite{fid} and CLIP Similarity (ClipSIM)~\cite{clipsim}, which assess the distance between image pair in feature space. 
To evaluate the subjective image quality, we use the Deep Image Structure and Texture Similarity
(DISTS)~\cite{dists} and no-reference Naturalness Image Quality Evaluator (NIQE)~\cite{niqe} to evaluate the perceptual quality.
%
%
Furthermore, we use bit per pixel (bpp) to evaluate the bitrate of the compressed images. 
%

%
\Paragraph{Implementation Details.}
We implement the proposed framework using Pytorch library. A pretrained SD 1.5 is implemented as the universal generator, with the guidance scale set to 7.5 and diffusion steps set to 50. The conditional scale of T2I-Adaptor is 1 for all priors. Layer-wise implementation details are listed below. (1) Semantic Layer: We use a pretrained BLIP-2 model as the encoder. The image captions are stored in and compressed using Zstd\footnote{\url{https://github.com/facebook/zstd}} with compression level 19. (2) Structure Layer: The edge detector utilizes a pretrained PiDiNet with threshold set to 50. Extracted edge maps are downsampled by a factor of 2 and compressed using VVC reference software VTM 15.0\footnote{\url{https://vcgit.hhi.fraunhofer.de/jvet/VVCSoftware_VTM/-/tree/VTM-15.0}} with QP 53. Detected poses are quantized to keep two decimal places and compressed using Zstd with compression level 19. (3) Texture Layer: Each color map is $8\times8$ with a bit depth of 8. The color codes are rendered into a color map before decoding. 

%

%
\vspace{-3mm}
\subsection{{Image Compression Performance}}
%
\Paragraph{Baselines.}
We compare our work with representative codes VVC~\cite{vvc}, JPEG2000~\cite{jpeg2000}, and PerCo~\cite{perco}. JPEG2000 is a classical codec for images and VVC is the latest hand-crafted codec for videos. Among neural compressors, PerCo is the state-of-the-art method utilizing the capability of stable diffusion to realize realistic image compression at the ultra-low bitrate.

\Paragraph{Qualitative Results.}
%
Fig.~\ref{fig:3-pose-qual} and Fig.~\ref{fig:2-edge-qual} provides qualitative comparisons of our work to baselines, with poses and edges as structures maps respectively. 
Our work yields the reconstructed images with more realistic details, higher semantic fidelity, and perceptual quality. JPEG2000 reconstructions are much blurry and VVC involves lots of blocking artifacts.
Perco achieves competitive visual qualities compared to JPEG2000 and VVC, while it is not able to generate controllable images with individual semantic, structure and texture conditioning, as our method does.

\begin{figure*}[!t]
    \centering
    \includegraphics[width=0.98\linewidth]{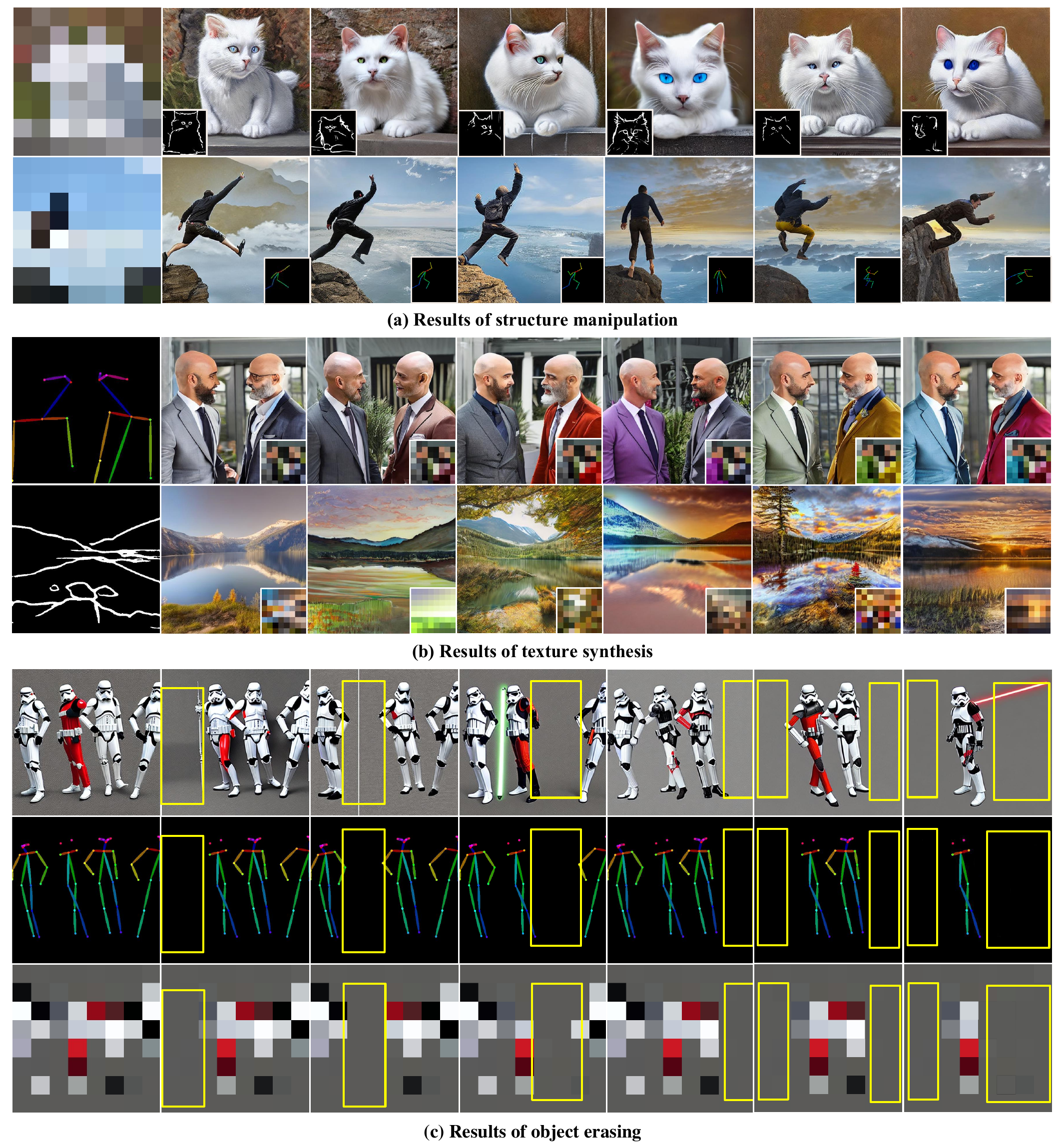} 
    \vspace{-3mm}
    \caption{Results of Structure Manipulation, Texture Synthesis and Object Erasing. The first column shows the unchanged texture/structure map, the remaining columns shows the editing results. The edited structure/texture map is plotted in the corner of each editing result respectively. }
    \label{fig:5-img-edit}
    \vspace{-5mm}
\end{figure*}

%
\Paragraph{Quantitative Results.}
As illustrated in Fig.~\ref{fig:4-quant}, our proposed method compresses images at ultra-low bitrates of less than 0.02 bpp while maintaining perceptual fidelity across a wide range of AIGC images. For all the metrics, our method significantly outperforms traditional codecs VVC and JPEG2000. For NIQE and FID, our method is almost on par with Perco. Meanwhile, the R-D curves of our method indicate that image fidelity improves as more structure and texture information is provided during decoding, demonstrating the scalability of our proposed multi-layered bitstream.


%

%
\vspace{-3mm}
\subsection{Image Editing Performance}
\label{sec:appl-img-edit}
%
\vspace{-2mm}

\Paragraph{Structure Manipulation.}
Demonstrative results of structure manipulation are shown in Fig.~\ref{fig:5-img-edit}. 
In the layered compression framework, modification of geometric structure can be achieved by directly modifying the structure map, e.g. changing the axes of keypoint, adding or subtracting objects. 
The decoder dynamically integrates semantic information and texture information into the updated structure map, synthesizing the edited image. 
%

%
\Paragraph{Texture Synthesis.}
Results of texture synthesis are shown in ~\figref{5-img-edit}. 
Different from many other compression methods that encodes texture into latent representation, we utilize the $8 \times 8$ color matrix as texture map, allowing a more intuitive and implicit editing process. 
The texture information can be update by changing parts of the color matrix or swapping texture maps across images. 
Results in \figref{5-img-edit} shows that our model has strong capacity of fitting texture representations into geometric structures and produce pleasant results. It indicates the potential of the proposed method to establish a connection between image compression and image manipulation.
%

%
\Paragraph{Object Erasing.}
We also show the results of object erasing in \figref{5-img-edit}. 
Unwanted objects can be wiped out by jointly masking out the areas in both structure and texture map, while the remain part of the image remain clear and natural. 
Object erasing is very useful in image editing and visual communication applications in real life. 
Moreover, the proposed method offers object erasing in a more efficient and simple way compared to manual methods via image editing software. 
The results reveals the potential of proposed method to combine image compression and manipulation. 
\vspace{-3mm}
\section{Conclusions}
\vspace{-2mm}
%
%
%
%
%
In this paper, we propose a layered cross-modal compression framework for AIGIs. We advocate that stable diffusion serves as a natural cross-modal decoder by leveraging rich and scalable priors. Specifically, our framework decomposes AIGIs into semantic, structure, and texture priors, effectively conveying multiple levels of visual information. 
The semantic prior functions as the foundational layer, and the structure and texture priors serve as enhancement layers.
Qualitative and quantitative results demonstrate the superiority of our proposed framework at ultra-low bitrates, outperforming VVC in both visual quality and objective metrics. Additionally, our framework enables seamless image editing through direct manipulation of the scalable bitstream, significantly facilitating downstream applications.

\Section{References}
\bibliographystyle{IEEEbib}
\bibliography{refs}

\end{document}

%% file: macro.tex
\usepackage{xspace}
\usepackage{color}
\usepackage{multirow}
\usepackage{ifthen}

\long\def\ignorethis#1{}

\definecolor{gray}{rgb}{0.35,0.35,0.35}
\definecolor{MyBlue}{rgb}{0,0.2,0.8}
\definecolor{MyRed}{rgb}{0.8,0.2,0}
\definecolor{MyGreen}{rgb}{0.0,0.5,0.1}
\definecolor{MyGray}{rgb}{0.4,0.4,0.4}

\newlength\paramargin
\newlength\figmargin
\newlength\subfigmargin
\newlength\secmargin
\newlength\subsecmargin
\newlength\tabmargin
\newlength\eqmargin

\usepackage{array}
\newcolumntype{L}[1]{>{\raggedright\let\newline\\\arraybackslash\hspace{0pt}}m{#1}}
\newcolumntype{C}[1]{>{\centering\let\newline\\\arraybackslash\hspace{0pt}}m{#1}}
\newcolumntype{R}[1]{>{\raggedleft\let\newline\\\arraybackslash\hspace{0pt}}m{#1}}

\def\eg{e.g.,~}

\newcommand{\secref}[1]{Section~\ref{sec:#1}}

\newcommand{\figref}[1]{Fig.~\ref{fig:#1}}

\newcommand{\Paragraph}[1]{\noindent\textbf{#1}}

%% file: main.bbl
\begin{thebibliography}{10}

\bibitem{sd}
Robin Rombach, Andreas Blattmann, Dominik Lorenz, Patrick Esser, and Bj{\"o}rn Ommer,
\newblock ``High-resolution image synthesis with latent diffusion models,''
\newblock in {\em Proceedings of the IEEE/CVF conference on computer vision and pattern recognition}, 2022, pp. 10684--10695.

\bibitem{controlnet}
Lvmin Zhang, Anyi Rao, and Maneesh Agrawala,
\newblock ``Adding conditional control to text-to-image diffusion models,'' 2023.

\bibitem{t2i}
Chong Mou, Xintao Wang, Liangbin Xie, Yanze Wu, Jian Zhang, Zhongang Qi, and Ying Shan,
\newblock ``T2i-adapter: Learning adapters to dig out more controllable ability for text-to-image diffusion models,''
\newblock in {\em Proceedings of the AAAI Conference on Artificial Intelligence}, 2024, vol.~38, pp. 4296--4304.

\bibitem{cmc}
Jiguo Li, Chuanmin Jia, Xinfeng Zhang, Siwei Ma, and Wen Gao,
\newblock ``Cross modal compression: Towards human-comprehensible semantic compression,''
\newblock in {\em Proceedings of the 29th ACM international conference on multimedia}, 2021, pp. 4230--4238.

\bibitem{vrcmc}
Junlong Gao, Jiguo Li, Chuanmin Jia, Shanshe Wang, Siwei Ma, and Wen Gao,
\newblock ``Cross modal compression with variable rate prompt,''
\newblock {\em IEEE Trans. on Multimedia}, 2023.

\bibitem{rdocmc}
Junlong Gao, Chuanmin Jia, Zhimeng Huang, Shanshe Wang, Siwei Ma, and Wen Gao,
\newblock ``Rate-distortion optimized cross modal compression with multiple domains,''
\newblock {\em IEEE Trans. on Circuits and Systems for Video Technology}, 2024.

\bibitem{rethink}
Pingping Zhang, Shiqi Wang, Meng Wang, Jiguo Li, Xu~Wang, and Sam Kwong,
\newblock ``Rethinking semantic image compression: Scalable representation with cross-modality transfer,''
\newblock {\em IEEE Trans. on Circuits and Systems for Video Technology}, vol. 33, no. 8, pp. 4441--4445, 2023.

\bibitem{gan}
Ian Goodfellow, Jean Pouget-Abadie, Mehdi Mirza, Bing Xu, David Warde-Farley, Sherjil Ozair, Aaron Courville, and Yoshua Bengio,
\newblock ``Generative adversarial networks,''
\newblock {\em Communications of the ACM}, vol. 63, no. 11, pp. 139--144, 2020.

\bibitem{perco}
Marlène Careil, Matthew~J. Muckley, Jakob Verbeek, and Stéphane Lathuilière,
\newblock ``Towards image compression with perfect realism at ultra-low bitrates,''
\newblock in {\em ICLR}, 2024.

\bibitem{misc}
Chunyi Li, Guo Lu, Donghui Feng, Haoning Wu, Zicheng Zhang, Xiaohong Liu, Guangtao Zhai, Weisi Lin, and Wenjun Zhang,
\newblock ``Misc: Ultra-low bitrate image semantic compression driven by large multimodal model,'' 2024.

\bibitem{text+sketch}
Eric Lei, Yiğit~Berkay Uslu, Hamed Hassani, and Shirin~Saeedi Bidokhti,
\newblock ``Text + sketch: Image compression at ultra low rates,'' 2023.

\bibitem{VISION}
David Marr,
\newblock {\em VISION: A Computational Investigation into the Human Representation and Processing of Visual Information},
\newblock MIT Press, 2010.

\bibitem{blip2}
Junnan Li, Dongxu Li, Silvio Savarese, and Steven Hoi,
\newblock ``Blip-2: Bootstrapping language-image pre-training with frozen image encoders and large language models,''
\newblock in {\em International conference on machine learning}. PMLR, 2023, pp. 19730--19742.

\bibitem{Zstd}
Y.~Collet and M.~Kucherawy,
\newblock ``Rfc 8878: Zstandard compression and the 'application/zstd' media type,'' 2021.

\bibitem{pidinet}
Zhuo Su, Wenzhe Liu, Zitong Yu, Dewen Hu, Qing Liao, Qi~Tian, Matti Pietik{\"a}inen, and Li~Liu,
\newblock ``Pixel difference networks for efficient edge detection,''
\newblock in {\em Proceedings of the IEEE/CVF international conference on computer vision}, 2021, pp. 5117--5127.

\bibitem{vvc}
Benjamin Bross, Ye-Kui Wang, Yan Ye, Shan Liu, Jianle Chen, Gary~J Sullivan, and Jens-Rainer Ohm,
\newblock ``Overview of the versatile video coding (vvc) standard and its applications,''
\newblock {\em IEEE Trans. on Circuits and Systems for Video Technology}, vol. 31, no. 10, pp. 3736--3764, 2021.

\bibitem{openpose}
Z.~{Cao}, G.~{Hidalgo Martinez}, T.~{Simon}, S.~{Wei}, and Y.~A. {Sheikh},
\newblock ``Openpose: Realtime multi-person 2d pose estimation using part affinity fields,''
\newblock {\em IEEE Trans. on Pattern Analysis and Machine Intelligence}, 2019.

\bibitem{1k21}
Jianhui Chang, Zhenghui Zhao, Lingbo Yang, Chuanmin Jia, Jian Zhang, and Siwei Ma,
\newblock ``Thousand to one: Semantic prior modeling for conceptual coding,''
\newblock in {\em 2021 IEEE International Conference on Multimedia and Expo (ICME)}, 2021, pp. 1--6.

\bibitem{humanbody}
Ruofan Wang, Qi~Mao, Shiqi Wang, Chuanmin Jia, Ronggang Wang, and Siwei Ma,
\newblock ``Disentangled visual representations for extreme human body video compression,''
\newblock in {\em 2022 IEEE International Conference on Multimedia and Expo (ICME)}, 2022, pp. 1--6.

\bibitem{structureandtexture}
Jianhui Chang, Zhenghui Zhao, Chuanmin Jia, Shiqi Wang, Lingbo Yang, Qi~Mao, Jian Zhang, and Siwei Ma,
\newblock ``Conceptual compression via deep structure and texture synthesis,''
\newblock {\em TIP}, vol. 31, pp. 2809--2823, 2022.

\bibitem{agiqa3k}
Chunyi Li, Zicheng Zhang, Haoning Wu, Wei Sun, Xiongkuo Min, Xiaohong Liu, Guangtao Zhai, and Weisi Lin,
\newblock ``Agiqa-3k: An open database for ai-generated image quality assessment,''
\newblock {\em IEEE Trans. on Circuits and Systems for Video Technology}, 2023.

\bibitem{fid}
Martin Heusel, Hubert Ramsauer, Thomas Unterthiner, Bernhard Nessler, and Sepp Hochreiter,
\newblock ``Gans trained by a two time-scale update rule converge to a local nash equilibrium,''
\newblock {\em Advances in neural information processing systems}, vol. 30, 2017.

\bibitem{clipsim}
Eric Lei, Yi\u{g}it~Berkay Uslu, Hamed Hassani, and Shirin~Saeedi Bidokhti,
\newblock ``Text+ sketch: Image compression at ultra low rates,''
\newblock in {\em ICML 2023 Workshop on Neural Compression: From Information Theory to Applications}, 2023.

\bibitem{dists}
Keyan Ding, Kede Ma, Shiqi Wang, and Eero~P Simoncelli,
\newblock ``Image quality assessment: Unifying structure and texture similarity,''
\newblock {\em IEEE Trans. on pattern analysis and machine intelligence}, vol. 44, no. 5, pp. 2567--2581, 2020.

\bibitem{niqe}
Anish Mittal, Rajiv Soundararajan, and Alan~C. Bovik,
\newblock ``Making a “completely blind” image quality analyzer,''
\newblock {\em IEEE Signal Processing Letters}, vol. 20, no. 3, pp. 209--212, 2013.

\bibitem{jpeg2000}
A.~Skodras, C.~Christopoulos, and T.~Ebrahimi,
\newblock ``The jpeg 2000 still image compression standard,''
\newblock {\em IEEE Signal Processing Magazine}, vol. 18, no. 5, pp. 36--58, 2001.

\end{thebibliography}
